\documentstyle{laa}
\begin{document} 
\thesaurus{} 
\title{Do Soft Gamma Repeaters Emit Gravitational Waves ?}
\author{J.A. de Freitas Pacheco}
\offprints{de Freitas Pacheco J.A.}
\institute{Observatoire de la C\^ote d'Azur, B.P. 4229, 
F-06304 Nice Cedex 4, France}
\date{Received date; accepted date} 
\maketitle
\begin{abstract}

Soft gamma repeaters are identified as highly magnetized
(B$\approx 10^{14}$ Gauss) neutron stars. Magnetic stresses
induce tectonic activity, and field
annihilation in faults is the ultimate energy source
for the observed $\gamma$-ray emission. As a consequence of 
the crustal cracking, the stored elastic energy
is converted into high  frequency (kHz) shear waves, that 
excite nonradial oscillation modes damped by gravitational wave
emission. This
class of objects should certainly be considered as potential
sources of gravitational waves that could be detected by
the present planned interferometric antennas like VIRGO or LIGO.

\end{abstract}

\section{Introduction}

Neutron stars are certainly one of the most popular potential sources of 
gravitational waves (GWs) (see, for instance, Bonazzola \& Marck 1994,
for a recent review on astrophysical sources of GWs).
 Rotating neutron stars  may have a time-varying
quadrupole moment and hence radiate GWs, by either having a tri-axial
shape or a misalignment between the symmetry and total angular momentum 
axes, which produces a wobble in the star motion. In the former case the
GW frequency is twice the rotation frequency, whereas in the latter two
modes are possible (Zimmerman 1980) : one in which the GWs have  the
same frequency as rotation, and another in which the GWs have twice the
rotation frequency (the first mode dominates by far at small wobble
angles while the importance of the second increases for larger values). The
merging together of two neutron stars is also a very attractive possibility
to produce GWs 
due to the huge energy power implied in the process, and the well known
dynamical behaviour of the system (Peters \& Mathews 1963). The
expected detection frequency of coalescence events by VIRGO
has recently been reviewed by de Freitas Pacheco (1997).

Another possibility for neutron stars radiate GWs is if they  oscillate
in nonradial modes, a mechanism discussed already in the late
sixties (Thorne \& Campolattaro 1967). Neutron
stars pulsating nonradially have
been neglected as potential sources of GWs in more recent studies. The 
reasons are the absence of a convincing mechanism to excite quadrupole
modes and the weak signal expected in the process.

In a recent paper, Cheng et al. (1995) call the attention to the
striking similarity, already noticed by past investigators, between
the statistical behaviour of soft $\gamma$-ray
repeaters (SGRs) and earthquakes. More than a hundred events
from the source SGR 1806-20 were already detected by the satellite
ICE, and from the analysis of such data, they have noticed the following
statistical properties that are also observed in
earthquakes: 1) power-law
energy distribution (Gutenberg \& Richter 1956); 2) log-symmetric
waiting time distribution (see also Hurley et al. 1994a); 3) a
robust correlation between waiting times of successive events;
 4) a weak or no correlation between amplitudes and waiting times.
The latter property distinguishes the behaviour of SGRs from
other X-ray bursters, that have burst energies nearly proportional to
the preceding waiting times (van Paradijs et al. 1988). If
the eruption of X-bursters is related to accretion, then the
reservoir must be replenished until the critical value is again
attained, a condition necessary for a new outburst.
 On the other hand, if the reservoir is the stored strain energy
in the crust of a young neutron star, this energy may not 
completely be depleted by quakes, which would explain the
absence of correlation between waiting times and the energy
of the event. In the case of the Earth, the cracking of the crust
occurs when a critical stress limit is reached. The quakes
generated by the cracking produce a relaxation of the stresses
to values below the threshold, and a series of events with
different intensities may occur, whose energy distribution
follows the Gutenberg-Richter power law. 

Neutron stars are believed to have solid crusts (but see
Haensel 1995)  and hence to develop a tectonic activity similar
to the Earth.
The dissipated elastic energy is sufficient to produce glitches in 
the rotation period of young pulsars and  to induce changes in the
surface magnetic field, caused by the appearance of moving
tectonic platelets (Ruderman 1991a, b; Burdyuzha et al. 1996, 1997).  

 If SGRs are indeed neutron stars with an important tectonic
 activity, we may further explore the analogy with earthquakes, 
considering that most of 
the strain energy goes into a 
mechanical channel, exciting  normal oscillating modes.
In this work the consequences of such a possibility are examined.
In the proposed scenario, the main features of the 
soft $\gamma$-ray emission may be explained, as a consequence of
 magnetic field annihilation in the faults produced during the
cracking of the crust.
The expected gravitational strain h$_0$ due to nonradial oscillations
excited by the release of elastic energy, is computed for
different neutron star models, and compared with the
 planned sensitivity of interferometric antennas like VIRGO.

\section{$\gamma$ and GW emission from SGR-like sources }

\subsection{The $\gamma$-ray emission}

Neutron stars are able to develop tectonic 
activity (see, for instance, Ruderman 1991c and references therein). 
Such an activity will probably depend on age, rotation
period and torques acting on the star, among other physical variables.
 Rotation discontinuities (glitches) are often observed
in pulsars with mean ages around 0.7-1.0 Myr (Shemar \& Lyne
1996) and, within 3 kpc from the sun, only about  2\% of the
pulsar population displays a glitch activity. On the
other hand, the number of SGRs
inside the Galaxy is unknown, and
presently only four sources are included in such a class of objects:
SGR 0526-66, SGR 1806-20, SGR 1900+14 and SGR 1814-13. The first was
suggested to be associated with the supernova remnant N49 in the
Large Magellanic Cloud, whereas the following two sources are
located in the galactic plane.  SGR 1806-20 may be 
associated to the supernova remnant G10.0-0.3, with an
angular distance of about $7^o$ from the galactic center. 
Distance estimates are quite uncertain, ranging
from 5 to 15 kpc (Wallyn
et al. 1995).  The last one was only recently proposed as a new member
of this class (IAUC 6743).
The small number of sources detected presently and the possible association
with supernova remnants suggest that the soft $\gamma$-ray
emission phase has
a short lifetime (10$^2$ - 10$^3$ years) and that the major
tectonic activity is strongly suppressed on such a timescale. The 
subsequent activity will produce quakes of lesser intensity,
which could be related to the glitch phenomena.

SGRs behave quite differently from other high energy transient
sources. Their typical photon energy is about 10-30 keV and the
burst characteristics are quite similar from one event to another.
In a first approximation, spectra can be represented by a black-body.
In this case, if the temperature is constrained within the above
interval, then the different observed values of the fluence require
variations of the emitting surface (Fenimore et al. 1994).

The scenario to be explored has essentially been developed
in the past years (Alpar et al. 1984, Ruderman 1991a). We assume that
neutron stars have a superfluid core and a solid crust with
a typical thickness {\bf L} of about  1 km. The neutron superfluid
fills the space between the crustal lattice nuclei, forming
a quasi-parallel array of quantized vortex lines, which are
expected to be pinned to those nuclei constituting the
crustal lattice (Alpar et al. 1984).

The solid crust as soon as formed, is under stresses
and the maximum shear stress that the crust
can bear before cracking is (Ruderman 1991b)
\begin{equation}
S_{max} \approx {{L}\over{R}}{\mu}{\theta_{max}} \approx 2\times
10^{26}({{\theta_{max}}\over{10^{-2}}}) \,\ erg.cm^{-3}
\end{equation} 
where R is the neutron star radius ($\approx$ 10 km), $\mu$ is
the lattice shear modulus and $\theta$ is the dimensionless strain.
The numerical value corresponds to a typical crust density of about
5$\times 10^{13}$ g.cm$^{-3}$. On the other hand, it is possible
that the stellar core protons form a type II superconductor and, in
this case, the core magnetic field would be organized into an array
of quantized flux tubes ending at the base of the crust. The
average field of these magnetic flux tubes is about
{\bf $B_q$} $\approx 10^{15}$ Gauss. The flux tubes do not 
interpenetrate the superfluid vortices and, as the core
vortex lines move out away from the spin axis, the magnetic
 flux tubes are forced to move with them (Srinivasan et al. 1990). The
 motion of these magnetic flux tubes generates
stresses on the base of the crust of the order of
\begin{equation}
S_{mag} \approx {{BB_q}\over{8\pi}}
\end{equation}
and cracking may occur if $S_{mag} \geq S_{max}$,  corresponding
 to a critical
crustal field {\bf B} greater than
$5\times 10^{12}({{\theta_{max}}\over{10^{-2}}})$ Gauss.
The critical field depends on the adopted value for the
strain $\theta_{max}$, a parameter rather uncertain in the
case of the crystal lattice that constitutes the crust
of neutron stars. The melting
temperature of the lattice is about 3$\times 10^9$ K, and
the actual temperature is probably much less. In this
case, assuming that laboratory experiments on Li crystals
can be extrapolated to the high pressures present
 in neutron star
crusts, $\theta_{max} \approx 10^{-2}$, although smaller
values have been suggested in the literature 
(Smoluchowski \& Welch 1970; Ruderman 1991b). Here, it
is assumed that the crust structure is characterized
by a $\theta_{max} = 10^{-2}$, implying that rather
high magnetic fields are required to produce significant
tectonic effects. In the present scenario, SGRs would be
associated to highly magnetized neutron stars. 
Similar scenarios have already been
invoked to explain the $\gamma$-ray burst phenomenon (Vietri 1996). 

As the magnetic stresses develop, the crust can be deformed
or cracked. In case
of cracking, magnetic field annihilation in the fault
reduces the local stresses, heats the medium and produces
a sudden increase in the local photon emission. Equating
 the Stefan energy flux to the
rate of magnetic energy per unit area dissipated in
the fault, a rough estimate of the resulting temperature
can be obtained 
\begin{equation}
T = 17.6{({{L}\over{1km}})^{1/4}}{({{1s}\over{t}})^{1/4}}
{({{B}\over{5\times 10^{12}G}})^{1/2}}{f_M^{1/4}} \,\ keV
\end{equation}
where $f_M$ is the fraction of the magnetic energy
dissipated in the event.

The total emitted photon energy corresponds essentially to
the field annihilation in the fault, namely
\begin{equation}
E_{\gamma} = 1.2\times 10^{42}f_S{({{B}\over{5\times 
10^{12}G}})^2}f_M \,\ erg
\end{equation}
where $f_S$ is the fractional area of the star surface corresponding to faults.  

In order to exemplify, let us assume a neutron star with
an initial field ${\bf B} = 10^{14}$ Gauss. If 10\% of
the initial magnetic energy is converted into heat, then from
eq. (3) the resulting temperature is about 
44 keV, consistent with the observed values. The total
emitted energy is 
$E_{\gamma} \approx 5\times 10^{39}({{f_S}\over{10^{-4}}})$
erg. A comparison with data requires the knowledge of the
source distance and the relative area of the faults. One
should emphasize that the $\gamma$-ray emission in our model
is {\bf not isotropic}, alleviating
energy requirements. We shall return to this problem later.

\subsection{The GW emission}

In the preceding section, we have shown that under certain
circumstances, the crust of a neutron star cracks due
to magnetic stresses. A fraction
of the initial magnetic energy is annihilated and
released as high energy photons. The stored elastic energy
is also released in the event, being converted into
shear vibrations with frequencies in the kHz regime
(Blaes et al. 1989). These waves are able
to excite nonradial modes, which will be damped by GW emission.

Considering the one-parameter approach of Baym \& Pines (1971), the
energy of a rotating neutron star with a solid crust is
\begin{equation}
E = E_{0} + {{1}\over{2}}I{\Omega^2}(1-\varepsilon) + W\varepsilon^2
+ B(\varepsilon_0 -\varepsilon)^2
\end{equation}
where $\varepsilon$ and $\varepsilon_0$ are respectively the actual stellar
oblateness and the reference oblateness, corresponding to the strain-free
situation; W measures the gravitational potential of the star and B measures
the stored elastic energy. If $\varepsilon \leq \theta_{max}$, oblateness
reducing crustquakes are possible. 
According to Pandharipande et al. (1976),  rotating neutron stars may
have B values around 
10$^{48}$ - 10$^{49}$ erg,
depending on the adopted equation of state. Thus, stars with initial rotation
periods of about 8 ms may have rotation induced stresses corresponding
to a stored elastic energy of about $10^{45}$ erg. 

Pulsar glitches may give another hint concerning the stored
elastic energy.
If glitches are associated with tectonic activity, then
 the observed
variation of the angular velocity is related to the released elastic
 energy by
(Baym \& Pines 1971)
\begin{equation}
\Delta E_{el} \approx W\theta_{max}{{\Delta \Omega}\over{\Omega}}
\end{equation}
where W is the gravitational potential energy of the star.

For the Crab pulsar ${{\Delta \Omega}\over {\Omega}} \approx$ 10$^{-8}$
and, from the above equation $\Delta E_{el} \approx$ 10$^{43}$ erg.
The glitches observed in the Vela pulsar correspond to relative variations
in angular velocity two orders of
magnitude higher, implying $\Delta E_{el} \approx$ 10$^{45}$ erg.
However, this theory predicts 
recurrence periods for glitches of the Vela pulsar of the order of  10$^3$ up
to 10$^4$ yr, which are few orders of magnitude higher than the observed value ($\approx$
2.8 yr).  This is because one assumes that 
the crust is under stresses  rotationaly induced, but if high
magnetic fields as discussed above, instead of rotation, are the origin of 
stresses, then crustquakes are still a possibility to explain 
macro-glitches, ( see, for instance, Blanford
1995). It is worth mentioning that the pulsar monitoring program at Jodrell
Bank (Shemar \& Lyne 1996), have evidentiated that nine in fourteen objects
display glitches with amplitudes ${{\Delta \Omega}\over{\Omega}} \approx$
10$^{-6}$, comparable to those observed in the Vela pulsar. Therefore, if such
an activity is tectonic in origin, values of about  10$^{45}$ erg for the energy
released are not unusual, if magnetic fields are indeed the main origin of
stresses.

Here we assume such a possibility, and we explore
the consequences of a given event  associated 
to an energy release  E = 10$^{45}$ erg, supposing that 
such an energy is essentially channeled
into nonradial pulsation modes.

Under these conditions, the expected signal-to-noise ratio
for a gravitational detector, if both
source and antenna have a quadrupole beam pattern is (Thorne 1987)
\begin{equation}
{({S\over N})^2} = 3<F^2>\int ^{\infty}_0{{<\mid \tilde h^2(\nu) \mid>}\over{S_h(\nu)}}d\nu
\end{equation}
where F is the detector beam pattern factor, $\tilde h(\nu)$ is the Fourier
transform of the gravitational strain amplitude and $S_h(\nu)$ is the noise
spectral density (in Hz$^{-1}$) of the detector. In the above equation, averages
are performed over the relevant angles defining the detector
geometry and the source orientation.

If we use the results by Thorne (1969), the gravitational strain
amplitude can be written as
\begin{equation}
h(t) = h_0e^{(i\omega_nt - {t\over\tau_n})}
\end{equation}
where $h_0$ is the initial amplitude, $\omega_n$ is the angular
frequency of the n-mode  and $\tau_n$ is the corresponding
damping timescale. The initial amplitude can easily be related to the
total energy E through the equation
\begin{equation}
{h_0} = 2{{({{GE}\over{c^3\tau_n}})}^{1/2}}{1\over{r\omega_n}}
\end{equation}
where G is the gravitational constant, c is the velocity of light and
r is the distance to the source.

The S/N ratio can now be derived calculating the Fourier transform
of equation (8) and substituting into equation (7). One obtains, after
integration
\begin{equation}
{S\over N} = 1.37\times h_0{{[{{\tau_n}\over{S_h(\nu_n)}}]}^{1/2}}
\end{equation}
The numerical factor was obtained using the explicit dependence of the
beam pattern F on angles appropriate to laser beam detectors
(Forward 1978; Rudenko \& Sazhin 1980).

If we impose $({{S}\over{ N}})$ = 3 (the confidence level usually required), equation (10) allows an estimate of $h_0$.
In fact, the choice of a specific
neutron star model and the sensitivity of the detector are also required, since the damping timescale and the noise
spectral density appear explicitly in equation (10).
The model is defined by adopting an equation of state, and by fixing the stellar mass. Once these parameters are established,  proper frequencies
and damping timescales can be computed. Concerning the noise spectral
density $S_h(\nu)$, we adopt the planned sensitivity for the interferometric
laser beam antenna VIRGO (A. Brillet 1997, private communication). 
In the present calculations, the pulsating properties of neutron star models as computed  by  Lindblom \& Detweiler (1983), are adopted. The expected
gravitational amplitude was computed for three different models based on
distinct equations of state, and  three different equilibrium configurations for
each model.  

The first column of table 1 gives the model number, following the
same notation as Lindblom \& Detweiler (1983). In particular, model N  is based
on Walecka's equation of state, whereas the others correspond to
softer equations of state. The second
column gives the stellar mass, and the three other  give respectively
the radius, the frequency
of the fundamental quadrupole mode and the corresponding damping
timescale. The resulting  gravitational strain amplitude
$h_0$ calculated from equation (10), which is essentially
a characteristic of the considered detector, is given in the sixth column of table 1.
We may now turn
to the source properties which affect the value of $h_0$, explicitly given
by equation (9). Since the fundamental frequency and the damping
timescale are determined by the model, the maximum
distances (in kpc) which may be probed by the detector in each
case are given in the last column of table 1.   

\begin{table}
\caption[]{Quadrupole Pulsational Properties of  Neutron Stars}
\begin{flushleft}
\begin{tabular}{lllllll}
\hline
Model&$M/M_{\odot}$&R(km)&$\nu_0(kHz)$&$\tau_0$(s)& 
$h_0$&$r_{max}$ \\
\hline
O &0.55&12.33& 1.36 & 1.30 & 8.9 $10^{-23}$&1.2 \\
O &1.57 &12.84& 1.72 & 0.22 & 2.7 $10^{-22}$&0.8 \\
O &2.38 &11.58  & 2.20 & 0.16&4.2 $10^{-22}$&0.5 \\
N &0.57&13.24 &1.25 & 1.55 & 7.5 $10^{-23}$&1.4 \\
N&1.45&13.82&1.51&0.30&2.0 $10^{-22}$&1.0 \\
N&2.56&12.27&2.08&0.18&3.7 $10^{-22}$&0.5 \\
M&0.49&17.46&0.88&4.67&3.3 $10^{-23}$&2.7 \\
M&1.44&15.79&1.33&0.40&1.6 $10^{-22}$&1.2 \\
M&1.76&11.91&2.24&0.14&4.6 $10^{-22}$&0.4 \\
\hline
\end{tabular}
\end{flushleft}
\end{table}

\section{Discussions}

In this work a scenario is proposed to explain the 
mechanism responsible for the observed $\gamma$-ray emission
from SGRs. The magnetic field topology in the crust, if the
neutron star has a superconducting core, may lead to field
annihilation between platelets or in the faults generated
by the cracking of the crust. Neutron stars with
initial magnetic fields of about 10$^{14}$ Gauss are required to
heat the medium to temperatures up 30-40 keV. However, the
present model predicts a total $\gamma$-ray emission not
higher than 10$^{39}$ erg, which is by far smaller than that
derived for the ''1979 March 5 '' event, produced by 
SGR 0526-66. If this source is associated with the supernova
remnant N49 in the LMC, then the energy released in soft
$\gamma$-rays would be about 10$^{44}$ erg. Such a huge amount
of energy is comparable, in our picture, to the total
amount of magnetic energy initially stored in the crust. This is
certainly a difficulty for the model (and for any other
model with a different energy reservoir), which can only
be relaxed if SGR 0526-66 is considerably nearby. It
should be emphasized that, by
different arguments, Fenimore et al. (1994) have also raised
some doubts about the localisation of SGR 0526-66 in the LMC. 

The amount of elastic energy stored before the 
cracking of the crustal lattice, is comparable to the
magnetic field energy
stored in the crust. Therefore such a reservoir of energy
would have the same difficulties to explain the ''March 5''
event, if the source would indeed be localized in the LMC.
Moreover, most of the elastic energy is converted into high
frequency shear waves, following the work by Blaes et
al. (1989), and these waves will likely excite the normal
oscillating modes of the star.

Under these conditions, the first point to be noticed is
that quadrupole modes will be damped by GW emission. The
frequencies of these modes
are in the kHz range, and they are included in the
wideband sensitivity curve of laser beam interferometers
like the french-italian antenna VIRGO, or the american
detector LIGO. The eventual detection of GWs from
these sources would represent an important tool for the
diagnosis of the neutron star interiors, since frequencies
and damping timescales depend on the equation of state, as
well as on the stellar mass. 
Inspection of table 1 shows that the oscillation frequency
 increases and the
damping time decreases with the mass of the
 star, irrespective of the
equation of state.  Low mass stars produce larger
 gravitational amplitudes
and can be seen more deeply in the Galaxy. The reason is that the duration
of the signal for those objects is longer than for high mass stars.

The present results indicate that distances up to 2.7 kpc
may be probed by the present planned sensitivity of VIRGO.
Nevertheless, if most of neutron stars have masses around
1.4 M$_{\odot}$ then the distances to be probed are of
the order of 1 kpc.
Moreover, the detection probability and the frequency of the events
depend on the badly known distribution of those sources
in the Galaxy. It is possible that there may be a large
number of SGRs to be discovered yet (Hurley et al. 1994a,b), since
the present experiments are probably missing sources with
either too long or
too short mean time intervals between bursts. 

SGRs should be considered as potential sources of GWs, and one of
the goals of this paper is to call the attention to such
a class of objects. It is worth mentioning that the
detection of GWs in coincidence with $\gamma$-photons may
represent an unambiguous possibility of
identifying a given class of sources, a desiderata for all GW experiments.

\vspace{0.5 cm}

{\it Acknowledgements: The author is grateful to M. de Boer and to
J. Horvath for their critical reading of the manuscript. The referee of
this paper, R. Epstein, is also thanked for his useful remarks.}

\vfill\newpage

\end{document}